\documentclass[12pt]{article}

\newcommand{\bpsi}{{\Psi}}
\newcommand{\bfi}{{\Phi}}

\newcommand{\dth}{\pa_\theta}
\newcommand{\tQ}{{\tilde Q}}
\newcommand{\hO}{{\hat\Omega}}
\newcommand{\hA}{{\hat A}}

\newcommand{\hd}{{\hat d}}
\newcommand{\hF}{{\hat F}}
\newcommand{\cas}{s}

\makeatletter
\@addtoreset{equation}{section}
\makeatother

\def\ftoday{{\sl {Le \number\day \space\ifcase\month 
\or janvier\or f\'evrier\or mars\or avril\or mai
\or juin\or juillet\or ao\^ut\or septembre\or octobre
\or novembre \or d\'ecembre\fi\space \number\year}}}    
\def\ptoday{{\sl {\number\day \space de\space \ifcase\month 
\or janeiro\or fevereiro\or mar{\c c}o\or abril\or maio
\or junho\or julho\or agosto\or setembro\or outubro
\or novembro \or dezembro\fi\space de\space \number\year}}}    
\def\gtoday{{\sl {Den \number\day. \ifcase\month 
\or Januar\or Februar\or M\"arz\or April\or Mai
\or Juni\or Juli\or August\or September\or Oktober
\or November \or Dezember\fi\space \number\year}}}    
\def\today{{\sl {\ifcase\month
\or January\or February\or March\or April\or May
\or June\or July\or August\or September\or October
\or November \or December\fi \space\number\day,\space 
                                            \number\year}}}

\renewcommand{\d}{\delta}         \newcommand{\D}{\Delta}

\newcommand{\m}{\mu}

\newcommand{\om}{\omega}         \newcommand{\OM}{\Omega}
\newcommand{\p}{\psi}              

\newcommand{\f}{{\phi}}           \newcommand{\F}{{\Phi}}

\newcommand{\HH}{{\cal H}}

\newcommand{\OO}{{\cal O}}
\newcommand{\PP}{{\cal P}}

\newcommand{\es}{\\[3mm]}

\newcommand{\sla}{\raise.15ex\hbox{$/$}\kern -.57em} 
\newcommand{\Sla}{\raise.15ex\hbox{$/$}\kern -.70em}

\newcommand{\lp}{\left(}\newcommand{\rp}{\right)}

\newcommand{\complex}{{\kern .1em {\raise .47ex
\hbox {$\scriptscriptstyle |$}}
    \kern -.4em {\rm C}}}
\newcommand{\real}{{{\rm I} \kern -.19em {\rm R}}}
\newcommand{\rational}{{\kern .1em {\raise .47ex
\hbox{$\scripscriptstyle |$}}
    \kern -.35em {\rm Q}}}
\renewcommand{\natural}{{\vrule height 1.6ex width
.05em depth 0ex \kern -.35em {\rm N}}}

\newcommand{\tr}{{\rm {Tr} \,}}

\newcommand{\pa}{\partial}

\newcommand{\dsum}[2]{\displaystyle{\sum_{#1}^{#2}}}   
\newcommand{\dint}{\displaystyle{\int}}

\newcommand{\twiddle}{\lower.9ex\rlap{$\kern -.1em\scriptstyle\sim$}}

\newcommand{\equ}[1]{(\ref{#1})}
\newcommand{\eq}{\begin{equation}}
\newcommand{\eqn}[1]{\label{#1}\end{equation}}
\newcommand{\eea}{\end{eqnarray}}
\newcommand{\eqa}{\begin{eqnarray}}
\newcommand{\eqan}[1]{\label{#1}\end{eqnarray}}
\newcommand{\ba}{\begin{array}}
\newcommand{\ea}{\end{array}}
\newcommand{\eqac}{\begin{equation}\begin{array}{rcl}}
\newcommand{\eqacn}[1]{\end{array}\label{#1}\end{equation}}

\newcommand{\bz}{\begin{enumerate}}
\newcommand{\ez}{\end{enumerate}}

\begin{document}


\hspace*{\fill}{{\normalsize 
\begin{tabular}{l}
{\sf hep-th/0303084} \\
{\sf LYCEN 2003-02}  \\
{\sf UFES-DF-OP2003/2}  \\
 {\sf \today} \\
 \end{tabular}   
 }}

\begin{center}
{\LARGE\bf Topological Yang-Mills Theories}\es
{\LARGE\bf and Their Observables:}\es
{\LARGE\bf a Superspace Approach\footnote{Talk presented by Olivier Piguet at
the 2nd International Londrina Winter School Mathematical Methods in
Physics, 2002, Londrina, Brazil.}}
\end{center}

\vspace{3mm}

\begin{center}{\large 
Jos\'e Luis Boldo$^{*}$,
Clisthenis P. Constantinidis$^{*,**}$, \es
Fran\c cois Gieres$^{***}$,
Matthieu Lefran\c cois$^{***}$
and Olivier Piguet$^{*}$
}\end{center}
\vspace{1mm}

\noindent $^{*}$ {\it Universidade Federal do Esp\'{\i}rito Santo 
(UFES), 
CCE, Departamento de F\'{\i}sica, Campus Universit\'ario
de Goiabeiras, BR-29060-900 - Vit\'oria - ES (Brasil)}

\noindent $^{**}$ {\it The Abdus Salam ICTP, Strada Costiera 11, 
I - 34014 - Trieste (Italy)}

\noindent $^{***}$ {\it 
Institut de Physique Nucl\'eaire,
Universit\'e Claude Bernard (Lyon 1), \\
43, boulevard du 11 novembre 1918,
      F - 69622 - Villeurbanne (France)}

{\tt E-mails: jboldo@cce.ufes.br,
clisthen@cce.ufes.br, \\
gieres@ipnl.in2p3.fr, 
lefrancois@ipnl.in2p3.fr, piguet@cce.ufes.br}

\vspace{8mm}

\begin{abstract}
Witten's observables of topological Yang-Mills theory,
defined as classes of an equivariant cohomology, are re-obtained 
as the
BRST cohomology classes of a superspace version of the theory.
\end{abstract}

\section{Introduction}
Observables in topological theories are global, such as knot invariants in
Chern-Simons theory, Wilson loops, etc. The problem of finding all of
them is a problem of ``equivariant cohomology'', as already pointed out 
by Witten in 1988 in the framework of 4-dimensional topological 
Yang-Mills theory\cite{witten,ouvry,delduc}. 
In the latter case, there exists an equivalent 
approach based on a superspace formalism  introduced  soon thereafter 
by Horne\cite{horne,carvalho}, 
which reduces the problem to that of looking for 
the BRST cohomology in a supersymmetric context. Our aim is to 
present and develop this formalism, as well as to generalize it 
to other models, such as topological gravity for instance.

This talk is the summary of a more extended work in 
preparation\cite{bcglp}.

\section{4-Dimensional Topological Yang-Mills Theory }

The gauge connection is given by the matrix valued 1-form
$a= a_\m^a T_a dx^\m$, where the matrices $T_a$ form a basis 
of the gauge Lie algebra, with
\eq
[T_a,T_b] = f_{ab}{}^c T_c\ ,\quad \tr (T_aT_b)= \d_{ab} \ .
\eqn{Lie-alg}

Topological Yang-Mills theory\cite{witten} is characterized,
beyond the
usual Yang-Mills gauge invariance, by the ``shift supersymmetry'' 
defined by the
infinitesimal transformations
\eq
\tQ  a = \p\ ,\quad \tQ \p = -d\f - [a,\f]\ ,\quad \tQ \f=0\ ,
\eqn{tilde-Q}
where the components of the 1-form $\p$ are fermions and $\f$ is a
bosonic 0-form. 

The infinitesimal gauge transformations 
parametrized by $\omega$ read as 
\[\ba{l}
\d_{\rm YM} a = d\om+[a,\om]\ ,\quad
\d_{\rm YM}\p= [\p,\om]\ ,\quad \d_{\rm YM}\f=[\f,\om]\ .
\ea\]
One checks that the square of the supersymmetry operator $\tQ$ is equal
to an infinitesimal gauge transformation, with parameter $\om$ 
substituted by the field $\f$.
Thus $\tQ$ is nilpotent on gauge invariant quantities. This led Witten
to define the {\it observables} of the theory as the cohomology classes of
$\tQ$ in the space of the gauge invariant operators, which is a problem of
{\it equivariant cohomology}\cite{ouvry,delduc}: a gauge invariant operator
$\OO$ belongs to the equivariant cohomology of $\tQ$ if
\[\ba{c}
\tQ \OO=0\ ,\qquad  \mbox{but}\quad \OO \not=\tQ\PP\ ,\quad 
 \mbox{with the conditions:}\quad 
   \d_{\rm YM}\OO=0\ ,\quad \d_{\rm YM}\PP=0\ .
\ea\]

One can also interpret this formalism as
the gauge fixing of the shift supersymmetry considered as a local
``gauge''
invariance, with a Yang-Mills-invariant gauge fixing condition of the type 
$f(a) \equiv  da + a^2=0$, or $f=*f$, where $*$ is the Hodge operator. 
Then, $\p$ is the ghost and $\f$ its ghost of ghost, $\tQ$ being the 
corresponding BRST operator.

\section{Superspace Formalism}

Enlarging the $d$-dimensional spacetime, of coordinates $x^\m\;
(\m=0,\cdots,d-1)$, with one fermionic dimension described by a 
Grassmann (i.e. anticommuting) coordinate $\theta$, with $\theta^2=0$,
we define a {\it superfield} as a superspace function 
\[
F(x,\theta) = f(x) + \theta f_{\theta}'(x)\ ,
\]
which transforms under a supersymmetry (SUSY) transformation as 
\eq
QF=\dth F\qquad\mbox{or, in components:}\quad Qf=f'\ ,\quad Qf'=0\ .
\eqn{SUSY-tr}
By construction the SUSY operator is strictly 
nilpotent:
$Q^2=0$.
Introducing the differential $d\theta$ -- a commuting quantity -- 
we may define {\it $p$-superforms},
\[
\hO_p (x,\theta) = \dsum{k=0}{p} \, \OM_{p-k}(x,\theta)(d\theta)^k \ ,
\]
where $\OM_{p-k}(x,\theta) = \om_{p-k}(x) + \theta\om_{p-k}'(x)$ is a
$(p-k)$-form in $x$-space with superfield coefficients, as well as
the {\it exterior superderivative:}
\eq
\hd = d+d\theta\dth 
= dx^\m\pa_\m+d\theta\dth\ ,\quad\mbox{with}\quad \hd^2=0\ .
\eqn{s-der}
A supergauge theory will be based on a {\it superconnection}, the
1-superform\footnote{All forms and superforms in this paper are Lie algebra valued: 
$\F=\F^aT_a$, see \equ{Lie-alg}.}
\[\ba{l}
\hA = A(x,\theta) + d\theta  \, A_\theta(x,\theta)
=  a(x)+\theta\p(x) +  d\theta\lp\chi(x)+\theta\f(x)\rp\ ,
\ea\]
a  Faddeev-Popov ghost 0-superform
\[
C (x,\theta)  =  c(x) + \theta  c'(x)\ ,
\]
and the corresponding {\it BRST transformations}
\[
\cas \hA = -( \hd C +[ \hA ,C ] )\quad  ,\quad 
\cas C =  -C^2\ , \qquad \cas^2=0\ .
\]
The latter read, in components:
\eq\ba{lll}
\cas a =  -Dc  \ ,\quad &\cas\p = - [c,\p]  -Dc'  \ ,\quad
                        &   \cas\chi =   - [c,\chi] - c'\ ,\\
\cas\f = - [c,\f] - [\chi,c'] \ ,\quad &\cas c  =  -c^2\ , \quad
 &  \cas c' =  -[c,c'] \ .
\ea\eqn{BRST-comp}


\subsection*{The Wess-Zumino gauge:}
%
One observes that the component field 
$\chi(x)$ is a pure gauge degree of freedom w.r.t. to
the $c'$-transformation in \equ{BRST-comp}. The gauge
invariance corresponding to the ghost $c'(x)$
can be fixed by the ``Wess-Zumino gauge'' condition
$\ \chi=0\;$.
This condition, which is not preserved by the SUSY transformations 
\equ{SUSY-tr}, is however stable with respect to 
the modified\cite{thompson} 
transformations
$\ \tQ = Q + \left. \cas \right| _{\chi =0, \, c=0,\, c'=\f}\ $
obtained by adding to the SUSY operator $Q$ a 
field dependent supergauge transformation. It is easy to check that this
new SUSY transformation reproduces,
for the ``physical'' fields $a$, $\p$ and $\f$,  the original 
transformations given by
\equ{tilde-Q}.

\subsection*{Cohomology of the SUSY operator $Q$:}
%
According to \equ{SUSY-tr}, the component fields are grouped in doublet 
representations of the nilpotent superspace SUSY operator $Q$.
It follows\cite{book,barnich} that 
{\it the cohomology of $Q$ is trivial}:
every form or superform which is $Q$-invariant is the $Q$-variation of
another form or superform. This obviously distinguishes $Q$ from its
Wess-Zumino gauge version $\tQ$ which is nilpotent only when applied
to gauge invariant quantities, and moreover has a nontrivial
equivariant cohomology\cite{witten}.


\section{Observables in the superspace formalism}

The equivalence of the superspace theory with the one originally 
proposed by Witten, the latter being a Wess-Zumino gauge fixed version
of the former, suggests to define an observable $\OO$ in superspace as
a BRST cohomology class:
\eq
\cas \OO = 0\ ,\quad\mbox{but}\quad \OO\not= \cas \PP\ ,
\eqn{s-cohom}
where $\OO$ and $\PP$ are both $p$-dimensional\footnote{The space-time
dimension will not be fixed a priori.
The space-time integrals are performed over 
an arbitrary $p$-manifold or submanifold $M_p$.}
space-time integrals:
\[
\OO=\dint_{M_p}\om_p(x)\ ,\quad \PP=\dint_{M_p}\f_p(x)\ ,
\]
with the conditions:
\eq
   Q\OO=0\ ,\quad Q\PP=0\ .
\eqn{SUSY-cond}
The latter condition and 
the triviality of the cohomology of $Q$ implies now that, up to a possible
total derivative which can be discarded without loss of generality, the
integrand of $\OO$ -- the $p-$form $\om_p$ -- is $Q$-exact:
\[
\om_p(x) = Q\OM_p(x,\theta)\ ,\quad\mbox{hence}\quad
\OO = \dint_{M_p}Q\OM_p(x,\theta) = \dint_{M_p}\dth\OM_p(x,\theta)\ ,
\]
which is the superspace integral of a ``superfield form'', i.e. of a
space-time form whose coefficients are superfields.

\section{General solution}

The general solution of the BRST cohomology problem \equ{s-cohom} with the 
SUSY inv\-ar\-ian\-ce condition \equ{SUSY-cond} consists of the two 
classes described below. For a proof that there are no other solutions, see
Reference \cite{bcglp}.

\subsection{Equivariantly Trivial Solutions} 
The first class of solutions is of the general form
\eq
\;^{D-p+1}\D_{(p)} = 
\int_{M_p} Q \;^{D-p}\HH_p(F_A,\bpsi,\bfi,{D_A}\bpsi,{D_A}\bfi)\ ,
\eqn{triv-obs}
where $\HH_p$ is a gauge invariant function of 
the superfield forms $F_A$, $\bpsi$ and $\bfi$ and their covariant
derivatives $D_A\bpsi$ = $d\bpsi+[A,\bpsi]$ and  $D_A\bfi$.
These superfield forms are the components of the supercurvature
$\hF \equiv \hd \hA + \hA^2$,  
\eq
\hF = {F_A} + \bpsi \, d\theta  + \bfi \, (d\theta)^2 \ , 
\eqn{s-curvature}
with 
\[\ba{l}
{F_A} = dA+A^2 \ , \quad
\bpsi = \partial_{\theta} A + D_A A_\theta 
= \psi +  D_a\chi + O(\theta)\ ,\\
\bfi  = \partial_{\theta} A_\theta + A_\theta ^2 
= \f+ \chi^2 + O(\theta)\ .
\ea\]
Their BRST transformations read as 
\[\ba{l}
\cas F_A =  - [ C,F_A]  \ ,\quad 
\cas\bpsi = - [C,\bpsi] \ ,\quad \cas\bfi = - [C,\bfi] \ .
\ea\] 
The solutions \equ{triv-obs}, although 
nontrivial in the sense of the
BRST cohomology, are trivial in the sense of the equivariant
cohomology. Indeed, 
in the Wess-Zumino gauge, they conserve the same form as in
\equ{triv-obs}, but with $A$, $F_A$, $\bpsi$ and $\bfi$ replaced by
$a$, $F_a=da+a^2$, $\p$ and $\f$, and the operator $Q$ replaced by $\tQ$:
the result is then explicitly given by the $\tQ$-variation of a gauge
invariant integral.

\subsection{Equivariantly Nontrivial Solutions} 

The solutions of the BRST cohomology 
problem \equ{s-cohom} 
which are nontrivial in the sense of the equivariant cohomology are
given in terms of superforms by the following superspace algorithm.
\vspace{3mm}

{\bf1.} Consider all the superforms $\hO_D(x,\theta)$ 
(of ghost number 0 and superform degree $D$) which are
nontrivial solutions of 
the cohomology $H(\cas|\hd)$
of $\cas$ modulo $\hd$ in the space of the superforms made
of polynomials 
of the basic superforms $\hA(x,\theta)$, $C(x,\theta)$, 
$\hd\hA(x,\theta)$ and $\hd C(x,\theta)$. Nontriviality in the sense of
the cohomology $H(\cas|\hd)$ for a superform $\hO$ means
\[
\cas\hO = 0\quad(\mbox{modulo\ }\hd)\ ,\qquad\mbox{but}\quad
\hO\not= \cas{\hat\F}\quad(\mbox{modulo\ }\hd)\ .
\]
These $\hO_D$ are nontrivial solutions of sets of ``superdescent equations''
\eq\ba{l}
\cas \hO_D + \hd\hO_{D-1}^1=0\ ,\quad
\cas \hO_{D-1}^1 + \hd\hO_{D-2}^2=0\ ,\quad 
\cdots \ ,\quad
\cas \hO_{0}^{D}=0 \ .
\ea\eqn{s-descent}
Here the exponent $g$ of the superform 
$\hO_p^g$ denotes its ghost number, i.e.
its degree in the superfield ghost $C$ or its components. The index $p$
denotes its degree as a superform.


\vspace{3mm}

{\bf 2.} Expand $Q\hO_D = \dth\hO_D$ according to the space-time 
form degree $p$:
\[
Q\hO_D = \dsum{p=0}{D} w_p(x) (d\theta)^{D-p} \ .
\]
The space-time forms $w_p$ are our solutions:
\[
sw_p(x) = 0 \quad(\mbox{modulo\ }d)\ ,\qquad\mbox{and}\quad
Qw_p (x) = 0 \ .
\]
This follows from the identity $\hd\,Q\,F(x,\theta)= d\,Q\,F(x,\theta)$,
which is a direct consequence of the definition \equ{s-der}.

We thus obtain, for a fixed maximum degree $D$, 
a set of observables
\[
\OO = \dint_{M_p} w_p(x) \ ,
\]
where the space-time forms $\om_p$ are the 
coefficients of the superform
$Q\hO_D$, where $\hO_D$ represents a nontrivial 
solution of the superdescent equations \equ{s-descent}.

\subsection{Solution of the Superdescent Equation}

The nontrivial observables are thus completely determined from the
general solution of the superdescent equations \equ{s-descent}. The
latter is given by the generalization to the present superspace
formalism of standard results of BRST cohomology\cite{barnich}. The
result is:
\eq
\hO_{D} = \theta^{\rm CS}_{r_1}(\hA) 
f_{r_2}(\hF)\cdots f_{r_L}(\hF)\ , \qquad 
\mbox{with}\quad  D= 2 \, \dsum{i=1}{L} \, m_{r_i} -1 
\ , \quad L \geq 1 \ ,
\eqn{3.xx}
where $f_r(\hF)$ is the supercurvature invariant of degree $m_r$ in 
$\hF$ corresponding to the gauge group Casimir operator of degree
$m_r$, and $\theta^{\rm CS}_{r}(\hA)$ is the associated Chern-Simons
form: 
\[
\hd \theta^{\rm CS}_{r}(\hA) = f_r(\hF)\ .
\]
We note that the superform degree of the solution \equ{3.xx} is odd.

One finally checks that, upon 
reducing the result to the Wess-Zumino gauge, one
recovers the observables originally given by Witten\cite{witten}.

\section{Example}

We consider the case of maximum degree $D=3$. The superdescent equations
read as
\[\ba{l}
\cas \hO_3 + \hd\hO_2^1=0\ ,\quad
\cas \hO_2^1 + \hd\hO_1^2=0\ ,\quad 
\cas \hO_1^2 + \hd\hO_0^3=0\ ,\quad
\cas \hO_0^3=0\ .
\ea\]
The unique nontrivial solution is 
\[\ba{l}
\hO_3 =  \tr ( \hA \hd\hA + \frac{2}{3} \hA^3 ) \ ,\quad
\hO_2^1 =  \tr (\hA \hd C ) \ ,\quad
\hO_1^2 =  \tr ( C \hd C ) \ ,\quad
\hO_0^3 =  -\frac{1}{3} \; \tr C^3 \ .
\ea\]
Note that $\hO_3$ is the Chern-Simons superform:
$\hd\hO_3 = \tr(\hF^2)$.

\vspace{2mm}

The observables are then given by the expansion
\[
\dth \hO_3 = \dsum{p=0}{3} w_p(x) (d\theta)^{3-p} \ ,
\]
with
\[\ba{ll}
w_0     =   \tr ( \f^2 +2\f\chi^2 ) \ ,\quad
&w_1      =  \tr 2 ( \p\f +\f D\chi + \p\chi^2 ) +d(\cdots)\\
w_2      = \tr ( 2\f F_a + 2\p D\chi +\p^2 )
+d(\cdots)\ ,\quad
&w_3      =     \tr (2\p F_a ) +d(\cdots)
\ea\]

In the Wess-Zumino gauge ($\chi=0$):
\[\ba{ll}
w_0     =   \tr ( \f^2)  \ ,\qquad
&w_1      =  \tr ( 2\p\f  ) +d(\cdots)\ ,\\
w_2      = \tr ( 2\f F_a + \p^2) +d(\cdots)\ ,\qquad
&w_3      =     \tr (2\p F_a ) +d(\cdots)
\ea\]

which corresponds to Witten's result up to total derivatives.

\section{Conclusion}

The superspace
BRST cohomology which we have advocated  as an alternative  definition 
of the observables of 
a topological theory of the Yang-Mills type, reproduces 
Witten's original result
using a rather straightforward extension 
to superspace of standard results on BRST cohomology\cite{barnich}.
However it
is also realized that another 
type of solutions exists, which are finally 
proved to be trivial in the sense of equivariant cohomology. 
The main difficulty 
which had to be overcome was to show\cite{bcglp}
that there are no other solutions 
to the problem of the superspace
BRST cohomology. These results lead us to wonder about the applicability 
of such an approach to
the construction of observables in more 
complex systems, for example
topological gravity and Yang-Mills theories 
with more than one supersymmetry generator. These are problems
under current investigation, and we hope to 
provide answers to these 
questions soon\cite{inprogress}.

\section*{Acknowledgments}

J.L.B., C.P.C. and O.P. thank the Conselho Nacional de
Desenvolvimento Cient\'\i fico e Tecnol\'ogico (CNPq -- Brazil) for
financial support.\\
C.P.C. also thanks the Coordena\c c\~ao de Aperfei\c coamento de 
Pessoal de N\'\i vel Superior (CAPES -- Brazil) for financial support,
and the Abdus Salam International Center for Theoretical Physics
(ICTP -- Italy) for hospitality during a three months stay
under its Associate Program.\\
O.P. acknowledges
the Institut de Physique Nucl\'eaire,
Universit\'e Claude Bernard (Lyon 1) for its 
kind hospitality during a
one month's stay as Professeur Invit\'e.


\end{document}